\documentstyle[12pt]{article}
\newcommand{\be}{\begin{equation}}
\newcommand{\ee}{\end{equation}}
\newcommand{\bea}{\begin{eqnarray}}
\newcommand{\eea}{\end{eqnarray}}

\begin{document}
%%%%%%%%%%%%%Title page%%%%%%%%%%%%%%%%%%%

\begin{center}
\begin{large}
{\bf  CFT on the Brane \\}
{\bf with a\\}
{\bf  Reissner-Nordstrom-de Sitter Twist \\}
\end{large}  
\end{center}
\vspace*{0.50cm}
\begin{center}
{\sl by\\}
\vspace*{1.00cm}
{\bf A.J.M. Medved\\}
\vspace*{1.00cm}
{\sl
Department of Physics and Theoretical Physics Institute\\
University of Alberta\\
Edmonton, Canada T6G-2J1\\
{[e-mail: amedved@phys.ualberta.ca]}}\\
\end{center}
\bigskip\noindent
\begin{center}
\begin{large}
{\bf
ABSTRACT
}
\end{large}
\end{center}
\vspace*{0.50cm}
\par
\noindent

We consider a brane universe in a  Reissner-Nordstrom-de Sitter
 background spacetime  of arbitrary dimensionality. It is shown
that the brane evolution is described by generalized Friedmann
equations  for radiative matter  along with a stiff-matter
contribution. On the basis of the (conjectured) dS/CFT correspondence,
we identify various thermodynamic properties of the brane.
It is then demonstrated that, when the brane crosses the
de Sitter cosmological horizon, the CFT thermodynamics
and Friedmann-like equations coincide. Moreover, the CFT entropy
is shown to be expressible in a generalized Cardy-Verlinde 
form. Finally, we  consider the  holographic entropy bounds
in this scenario.

%PACS 04.70.Dy
\newpage

\section{Introduction}
%\medskip
\par

The so-called ``holographic principle'' \cite{tho,sus} has had considerable
influence on recent and current gravitational physics.  
The premise  of  this principle, roughly speaking, is that the maximal entropy
in any given volume will  be determined by the largest black
hole that fits inside of that volume.  On this basis, it
has been argued that all  relevant degrees of freedom 
of any given system  must, in some sense, ``live'' on  the boundary of
the system.
\par
One of the more prolific manifestations of the holographic principle
has been the  AdS/CFT correspondence \cite{mal,gub,wit}. 
More specifically, it has been  convincingly argued  that 
the thermodynamics at the horizon  of a  $n$+2-dimensional anti de-Sitter
(AdS) black hole can be  identified with  a certain $n$+1-dimensional
conformal field theory (CFT).  This dual CFT is assumed
to be a strongly-coupled one  and to live on a
timelike surface that can be identified as an asymptotic boundary
of the AdS spacetime.
\par
In a  breakthrough paper \cite{ver}, Verlinde  directly applied
the above ideas to an $n$+1-dimensional, radiation-dominated, closed 
Friedmann-Robertson-Walker (FRW) universe.\footnote{For earlier
studies on holography in a cosmological context, see Refs.\cite{first}-
\cite{last}.} 
This paper covered much ground, but the following two discoveries are of
particular interest. (i)  The AdS/CFT correspondence
 leads to an entropy that can be
expressed in terms of a generalized Cardy formula \cite{car}; with  
 the Cardy central charge being  directly related to the Casimir
energy.\footnote{In this context, the Casimir energy
refers specifically to the sub-extensive contribution to
the thermal energy. Furthemore, we will refer to the corresponding entropy
as the ``Casimir entropy''.} (ii) When the ``Casimir entropy''  
 saturates a certain bound (identifiable
with the holographic Bekenstein-Hawking entropy \cite{bek,haw}
 of a universal-size
black hole), then  the  Friedmann equations coincide precisely
with the generalized Cardy formula. To express this more elegantly,
the CFT and FRW equations remarkably merge at a  holographic
saturation point; thus implying that both sets of
equations arise from some fundamental, underlying theory.
\par
In Ref.\cite{sav}, Savonije and Verlinde  have extended these
notions to the case of a  Randall-Sundrum brane world \cite{ran,cos}
in the background of an AdS-Schwarzschild (black hole) geometry. 
In this scenario,
the $n$+1-dimensional CFT is regarded as living on the brane, which serves
as a boundary for the $n$+2-dimensional AdS bulk spacetime.
Given  a suitable choice of boundary conditions,  Savonije and Verlinde have
 shown
that the brane world corresponds to a FRW universe, with the brane dynamics
being described by the Friedmann equations for radiative matter.
Moreover, they demonstrated that the CFT thermodynamic relations
coincide with the Friedmann  equations  at the point when the
brane crosses the black hole (outermost) horizon. 
\par
Let us note that the  Verlinde-Savonije  treatment   has
since  been extended and generalized in a large array of
studies \cite{first2}-\cite{last2}.
\par
In analogy to the highly successful AdS/CFT duality, a dS/CFT
correspondence has also been hypothesized \cite{str}.\footnote{For
earlier works in this regard, see Refs.\cite{first3}-\cite{last3}.}
Naively, any  de Sitter (dS) spacetime can be   related to an AdS space   
via a simple
sign change in the cosmological
constant (negative for AdS and positive for dS). However,
 there are dire implications
 that make the proposed dS/CFT duality a significantly more challenging 
prospect. For instance,  dS spacetime lacks a 
 globally timelike Killing vector and a spatial infinity, while
  the black hole horizon (and  
related properties)
 have an
ambiguous observer
dependence.
 (For a comprehensive discussion on
dS spacetimes, see Ref.\cite{str2}.)  On  a more fundamental level,
  dS solutions are conspicuously absent in  string theories
(and  other forms of quantum gravity); thus  inhibiting
any rigorous  test of the proposed duality.
\par
In spite of the inherent  complications,  
there has still been substantial progress 
towards a holographic understanding of dS space 
\cite{str}-\cite{new3}.  
    With respect to the proposed dS/CFT 
correspondence, the higher-dimensional horizon is taken to be the
  dS cosmological horizon, while
the dual CFT is regarded as a Euclidean one that lives on a spacelike
asymptotic boundary.  
Essential to this construction is the existence of a certain duality: 
  time evolution in the dS bulk 
with  the reverse of a   renormalization group (RG) flow. Significantly,
this RG flow occurs between Euclidean CFTs
at past and future infinity \cite{str3}. 
\par
Very recently,  work has begun on adapting the  Verlinde-Savonije
program \cite{ver,sav} to a dS holographic picture \cite{new1}-\cite{new3}.
Given that recent observational evidence points to
us living in a dS universe \cite{obs}, the importance of
such studies probably cannot be overstated. On this note, 
the purpose of the current
paper is to see if the outcomes of Ref.\cite{sav} hold up
for a yet-to-be-considered scenario. Namely,  a brane universe
in a  Reissner-Nordstrom-de Sitter (RNdS) background.\footnote{By
incorporating charge into the model, we are following the
work of Refs.\cite{bis,myux,cai2}
 and  (in particular) 
Ref.\cite{you3}; all of which 
considered a charged AdS background.}
 
\par
The rest of the paper proceeds as follows.  In Section 2,
we identify the thermodynamics properties of the RNdS cosmological
horizon. This is followed by a formulation of the brane dynamics,
which leads to Friedmann-like cosmological  equations.
In Section 3, we apply the dS/CFT correspondence to
obtain the thermodynamics of a Euclidean CFT that lives
on the brane. Also,  the (generalized) Friedmann equations are
rewritten  in a form for which their connection with CFT thermodynamics
is manifest.
In Section 4, we consider  when the brane crosses
the cosmological horizon and consequently demonstrate
that the CFT thermodynamic properties and  Friedmann equations coincide
at precisely  this point.  Furthermore, the CFT entropy
is shown to  be expressible in a generalized Cardy-Verlinde \cite{car,ver}
form. Section 5   considers the implications of the prior analysis
with regard to holographic entropy bounds. Finally, Section 6 closes with
a summary and brief discussion.

\section{Bulk Thermodynamics and Brane Cosmology}
\par
Let us begin by considering an $n$+1-dimensional brane of constant tension
in an $n$+2-dimensional Reissner-Nordstrom-de Sitter (RNdS) background.
In a suitably static gauge, the bulk solution can be written
as follows \cite{man}:
\be
ds^2_{n+2}=-h(a)dt^2+{1\over h(a)}da^2+a^2d\Omega^2_{n},
\label{1}
\ee
\be
h(a)=1-{a^2\over L^2}-{\omega_{n+1}M\over a^{n-1}}+{n\omega_{n+1}^2
Q^2\over 8(n-1)a^{2n-2}},
\label{2}
\ee
\be
\omega_{n+1}={16\pi G_{n+2}\over n V_n},
\label{3}
\ee
\be
\phi_{dS}(a)={n\over 4(n-1)}{\omega_{n+1}Q\over a^{n-1}}.
\label{4}
\ee
Here, $L$ is the curvature radius of the dS background, $d\Omega_n^2$
is a unit $n$-sphere with volume $V_{n}$, $G_{n+2}$ is
the $n$+2-dimensional Newton constant, $M$ and $Q$
represent the conserved quantities of black hole mass and charge,
and $\phi_{dS}(a)$ is a measure of  the electrostatic potential
at $a$.\footnote{Since
there is no spatial infinity in RNdS space to use as a reference point,
$\phi_{dS}$ should not be interpreted as a literal electrostatic potential, as
it was in  analogous  AdS studies \cite{bis,cai2,you3}.}  
\par
In any dS space, there is a well-defined cosmological horizon having
similar thermodynamic properties to that of a black hole horizon
\cite{str2}. In our case, this cosmological horizon ($a=a_H$)  
corresponds to the largest root of $h(a)=0$. Thus, the following
useful relation can be obtained:
\be
1-{a_{H}^2\over L^2} -{\omega_{n+1}M\over a_H^{n-1}}+{n\omega_{n+1}^2
Q^2\over 8(n-1)a_H^{2n-2}}=0.
\label{5}
\ee
\par
In analogy with black hole thermodynamics \cite{gh}, the cosmological
horizon has an associated temperature, entropy and chemical potential
that are respectively given as follows:\footnote{In particular, the inverse
temperature can be identified with the periodicity of Euclidean
time and the entropy  with one quarter of the horizon surface area \cite{gh}.
The chemical potential has been defined in analogy to the
RN-AdS case \cite{you3}.}
\be
T_{dS}={(n+1)a_{H}^2-(n-1)L^2\over 4\pi L^2 a_H} +{n\omega_{n+1}^2 Q^2
\over 32\pi a_H^{2n-1}},
\label{6}
\ee
\be
S_{dS}={a_H^nV_n\over 4 G_{n+2}},
\label{7}
\ee
\be
\Phi_{dS}=\phi_{dS}(a_H)={n\over 4(n-1)}{\omega_{n+1}Q\over a_H^{n-1}}.
\label{8}
\ee 
The premise of the dS/CFT correspondence is that the above thermodynamics
can be identified, in some appropriate manner, with a Euclidean CFT
that resides (in the RNdS space) 
on a spacelike boundary at temporal infinity. We will
have more to say on this later.
\par
Let us now consider the brane, which can be regarded as a dynamical
boundary of the RNdS geometry. The brane dynamics are to be described
here via the following boundary action:
\be
{\cal I}_b={1\over 8\pi G_{n+2}}\int _{\partial {\cal M}}
\sqrt{\left| g^{ind}\right|}{\cal K}
+{\sigma\over 8\pi G_{n+2}}\int_{\partial {\cal M}}
\sqrt{\left| g^{ind}\right|},
\label{9}
\ee
where $g^{ind}_{ij}$ is the induced metric on the boundary  
($\partial{\cal  M}$),
${\cal K}\equiv {\cal K}^i_i$ is the trace of the extrinsic curvature
and $\sigma$ is a parameter measuring the brane tension.
Varying this action with respect to the induced metric, one
obtains (assuming a single-sided brane scenario) 
the following equation of motion:
\be
{\cal K}_{ij}={\sigma\over n}g_{ij}^{ind}.
\label{10}
\ee
\par
Following Ref.\cite{sav}, we can clarify the brane dynamics by introducing
a new (Euclidean) time parameter, $\tau$; whereby  $a=a(\tau)$,
$t=t(\tau)$ and:
\be
{1\over h(a)}\left({da\over d\tau}\right)^2-h(a)\left({dt\over
d\tau}\right)^2=1.
\label{11}
\ee
Unlike in the original analysis \cite{sav}, we have chosen $\tau$
so that the resulting line element is spacelike. This choice
naturally reflects the (presumed)  duality of a dS spacetime
 with a Euclidean CFT \cite{str}.
\par
Substituting the above condition into Eq.(\ref{1}), we see
that the induced brane metric takes on a Euclidean FRW form.
That is:
\be
ds^2_{n+1}=d\tau^2+a^2(\tau)d\Omega^2_{n}.
\label{12}
\ee
With this formulation, it is  clear that the radial distance, $a=a(\tau)$,  
measures the size of the
 $n$+1-dimensional brane universe.
\par
Let us now return considerations to the equation of motion (\ref{10}).
The extrinsic  curvature can  readily be  calculated (see, for instance, 
Ref.\cite{chr}) and then expressed in terms of the functions
$a(\tau)$ and $t(\tau)$. For any of the angular
components of the induced metric, this process yields:
\be
{dt\over d\tau}={\sigma a\over n h(a)}.
\label{13}
\ee
\par
Next, we define the Hubble parameter  in the usual way:
$H\equiv {\dot a}/a$.\footnote{Dots will always denote
differentiation with respect to the cosmological time parameter, $\tau$.}
Then Eq.(\ref{11}) can be re-expressed as follows: 
\be
H^2={1\over a^2}-{1\over L^2} -{\omega_{n+1}M\over a^{n+1}}+{n\omega_{n+1}^2
Q^2\over 8(n-1)a^{2n}}+{\sigma^2\over n^2},
\label{14}
\ee
where Eqs.(\ref{2},\ref{13}) have also been incorporated.
We are free to set $\sigma^2=n^2/L^2$ and conveniently cancel off
the $a$-independent terms. This choice yields a generalized
(first) Friedmann equation:
\be
H^2={1\over a^2}- {\omega_{n+1}M\over a^{n+1}}+{n\omega_{n+1}^2
Q^2\over 8(n-1)a^{2n}}.
\label{15}
\ee
\par
Furthermore, we can take the $\tau$ derivative of the above equation to obtain
the corresponding  second Friedmann equation:
\be
{\dot H}= -{1\over a^2}+ 
{(n+1)\omega_{n+1}M\over 2  a^{n+1}}-{n^2\omega_{n+1}^2
Q^2\over 8(n-1)a^{2n}}.
\label{16}
\ee
Note that the bulk RNdS background effectively induces both radiative matter
($\sim M/a^{n+1}$) and stiff matter ($\sim Q^2/a^{2n}$)
in the brane universe \cite{bis}.

\section{Euclidean CFT on the Brane}

Let us re-establish the premise of the dS/CFT correspondence
as it applies to our  model.  The  thermodynamic
properties of the cosmological horizon can presumably be identified
with the thermodynamics of a dual CFT that is Euclidean and 
living on the brane.
In analogy to the AdS analysis  of Savonije and Verlinde \cite{sav},
we will use this proposed duality  in defining 
the brane (CFT) thermodynamics. 
 Although the  process
is not so well defined for  dS spacetimes, we can proceed
(at least naively) by making some conjectural identifications.
\par
We begin here by applying the knowledge that the metric
for a boundary CFT can only be determined up to
a conformal factor \cite{gub,wit}. With this in mind,  let us 
  consider the asymptotic form of the Euclidean RNdS metric:
\be
\lim_{a\rightarrow\infty}\left[{L^2\over a^2}ds^2_{n+2}\right]=
dt_{E}^2+L^2d\Omega^2_n,
\label{17}
\ee
which can also be identified with the line element for
the relevant Euclidean CFT.
  Evidently, the Euclidean CFT time 
 ($t_E$) must be scaled by a factor of $a/L$
if the radius of the spatial sphere is to
 be set  equal to $a$. Proceeding on this 
basis, we  deduce that the same factor $a/L$ will
appear in the various expressions which relate the
thermodynamic properties of the dual spacetimes. (With one notable
exception being the entropy \cite{wit}.)
\par
Given the above consideration,  
we can  appropriately  identify
the thermodynamic properties of the CFT as follows \cite{sav}:
\be
E\equiv E_{CFT}=-{LM\over a},
\label{18}
\ee
\bea
T\equiv T_{CFT}&=&{L\over a}T_{dS}
\nonumber \\
&=& {1\over 4\pi a}\left[ {(n+1)a_{H}\over L}-{(n-1)L\over  a_H} +
{n L\omega_{n+1}^2 Q^2 
\over 8 a_H^{2n-1}}\right],
\label{19}
\eea
\bea
S\equiv S_{CFT}&=&S_{dS} 
\nonumber \\
&=& {a_{H}^n V_{n}\over 4 G_{n+2}},
\label{20}
\eea
\bea
\Phi\equiv \Phi_{CFT}&=&-{L\over a}\Phi_{dS}
\nonumber \\
&=&-{nL\over 4(n-1)}{\omega_{n+1}Q\over a a_H^{n-1}}.
\label{21}
\eea
\par
A commentary regarding the negative sign in Eq.(\ref{18})
(as well as Eq.(\ref{21})) is in order. It has recently been 
demonstrated that a dS black hole represents an excitation of
negative energy \cite{baln}. This counter-intuitive result
can be attributed to the unusual binding interaction that arises
between a  positive-mass object and a dS gravitational
field \cite{myun}. With this observation in mind, we have
followed prior works \cite{new1,new3} and identified the gravitational
energy with the negative of the mass observable.\footnote{Note
that, in Eq.(\ref{18}), we have omitted the energy associated with
the pure (i.e., $M=0$) dS background. This is consistent with 
 the convention initiated in Refs.\cite{ver,sav}.}
\par
In similar fashion to the energy, we have reversed the sign of the 
CFT chemical potential (\ref{21}).
 This choice can be further justified by noting that
a rotation to the Euclidean sector typically
requires an accompanying complexification of charge
\cite{gh}. Hence, we are inclined to transform $Q^2\rightarrow -Q^2$,
which is effectively the same as introducing a negative sign in Eq.(\ref{21}).
\par
Let us now suitably define an energy density ($\rho\equiv E/V$), pressure
($p\equiv \rho/n$)\footnote{Note that $p=\rho/n$ is the standard
equation of state for  radiative matter.}, electrostatic potential  
($\phi\equiv L\phi_{dS}/a$) and charge density 
($\rho_Q\equiv Q/V$); where
$V=a^nV_n$. With these definitions, the first and second 
(generalized) Friedmann 
equations (\ref{15},\ref{16})
can now be written as follows:
\be
H^2={16\pi G\over n(n-1)}\left[\rho-{1\over 2}\phi\rho_Q\right]
+{1\over a^2},
\label{22}
\ee
\be
{\dot H}=-{8\pi G\over (n-1)}\left[\rho+p-\phi\rho_Q\right]-{1\over a^2},
\label{23}
\ee
where $G=(n-1)G_{n+2}/L$ is the effective Newton constant on the
brane.\footnote{This  relation between bulk and brane gravitational constants
is the usual one for a Randall-Sundrum brane world (generalized
to arbitrary dimensionality) \cite{ran}.}
Notably, the cosmological evolution
has been directly   related 
to the energy density and pressure of radiative matter,
along with an electrostatic  energy density. The latter
 can be interpreted
as a stiff-matter contribution from a brane perspective \cite{bis}.
\par
It is interesting (and useful later on) to note that the Friedmann
equations can also be expressed  in   following suggestive forms:
\be
S_H={2\pi a\over n}\sqrt{-E_{BH}\left[2(E-{1\over 2}\phi Q)-E_{BH}\right]},
\label{24}
\ee
\be
E_{BH}=n\left[E+pV-\phi Q-T_HS_H\right],
\label{25}
\ee
where we have defined:
\be
S_{H}\equiv (n-1) {HV\over 4G},
\label{26}
\ee
\be
E_{BH}\equiv-n(n-1){V\over 8\pi G a^2},
\label{27}
\ee
\be
T_{H}\equiv -{{\dot H}\over 2\pi H}.
\label{28}
\ee
\par
The first Friedmann equation (Eq.(\ref{22}) or (\ref{24})) can 
further be expressed in the following intriguing manner:
\be
S_{H}^2=-2(S_B-S_Q)S_{BH}+S_{BH}^2,
\label{29}
\ee
where:
\be
S_{B}\equiv -{2\pi a\over n} E,
\label{30}
\ee
\be
S_{BH}\equiv {(n-1)\over 4 G a} V,
\label{31}
\ee
\be
S_Q \equiv   -{2\pi a\over n}\cdot{1\over 2}\phi Q.
\label{32}
\ee
\par
The parameters of
Eqs.(\ref{26}-\ref{28},\ref{30}-\ref{31})
are identically defined (up to a sign when appropriate) to those 
found in  Ref.\cite{ver}.\footnote{We also
 point out that $S_Q$ of Eq.(\ref{32})
is identically defined (up to the sign) with the parameter found in
Ref.\cite{you3}.} 
 For an AdS bulk, each of 
these parameters
significantly plays a role with regard to holographic bounds. 
(See Refs.\cite{ver,
sav} for a complete discussion.) Here,  we have introduced
the parameters  for illustrative   convenience and point out that
 their respective roles  do  not necessarily
persist in a dS holographic theory.  
We consider this further in Section 5.
\par
It should be kept in mind that 
%assuming an expanding brane universe (i.e., $H>0$),  
$S_{H}^2$, $S_{B}$, $S_{BH}$ and $S_Q$
are all strictly non-negative. Hence, given the
unorthodox form of Eq.(\ref{29}), one might anticipate
that severe cosmological constraints must be imposed
on the brane dynamics for a dS bulk spacetime. Again,
we defer this discussion until Section 5.

\section{Thermodynamics at the Horizon and the Cardy-Verlinde Entropy}

As was remarkably demonstrated in Ref.\cite{sav} for a bulk
AdS spacetime, the CFT thermodynamic relations  coincide
 with the Friedmann equations
when the brane crosses the black hole horizon.
We will now endeavor to see if the same behavior occurs at the
cosmological horizon of a bulk RNdS spacetime.
\par
First, let us compare Eq.(\ref{15}) for $H^2$ with  the equation
that defines the cosmological horizon (\ref{5}). A brief
inspection reveals that  the Hubble constant 
must obey:
\be
H=\pm {1\over L}\quad\quad\quad at\quad a=a_{H}.
\label{33}
\ee
The $+$ sign indicates an expanding brane universe, while
the $-$ sign describes a brane universe that is contracting.
For the sake of simplicity, we will focus on the expanding ($H=+L^{-1}$)
case.
\par
Next, let us reconsider the CFT entropy (\ref{20}). Evidently
(and as expected via the second law of thermodynamics),
this total entropy remains constant as time varies.
However, this is not the case for the entropy density:
\be
s\equiv {S\over V}= {(n-1)a_H^n\over 4G L a^n},
\label{34}
\ee 
which clearly evolves along  with the brane radius.
When the brane crosses the horizon, this entropy density takes on the form:
\be
s={(n-1)H \over 4G}\quad\quad\quad at \quad a=a_H,
\label{35}
\ee
from which it follows that (cf. Eq.(\ref{26})):
\be
S=S_H \quad\quad\quad at\quad a=a_H.
\label{36}
\ee
\par
It is also  instructive to consider the CFT temperature (\ref{19})
at the horizon-brane coincidence point. Using Eq.(\ref{23}) for ${\dot H}$,
as well as Eqs.(\ref{5},\ref{28},\ref{33}), we obtain the following:
\be
T=-{{\dot H}\over 2\pi H}=T_H \quad\quad\quad at\quad a=a_H.
\label{37}
\ee
Hence, when the brane crosses the horizon, the CFT entropy and 
temperature take on forms
that are simply expressed in terms of the Hubble parameter
(and its derivative).  These expressions are universal
inasmuch as they do {\bf not} depend explicitly
on  the  mass ($M$) and charge ($Q$) of the RNdS solution.
\par
Let us now introduce a quantity that can be readily identified
with the Casimir energy \cite{ver,sav}:
\be
E_C\equiv n\left[E+pV-\Phi Q-TS\right].
\label{38}
\ee
Given that $T=T_H$, $S=S_H$ and $\Phi=\phi$ at the coincidence point,
it follows that (cf. Eq.(\ref{25})):
\be
E_C=n\left[E+pV-\phi Q-T_H S_H\right]=E_{BH} \quad\quad\quad at \quad a=a_H.
\label{39}
\ee
We elaborate on the significance of this Casimir energy below.
\par
Let us now return to   the case of a general brane radius.
It can readily be verified that the CFT thermodynamic properties
satisfy the usual first law of thermodynamics. That is:
\be
TdS=dE-\Phi dQ+PdV.
\label{40}
\ee
Much can be revealed if we re-express the first law in terms of densities.
In this case:
\be
Tds=d\rho-\Phi d\rho_Q +n\left[\rho+p-\Phi\rho_Q-Ts\right]{da\over a},
\label{41}
\ee
where we have applied the equation of state ($p=\rho/n$)
and $dV=nVda/a$ (since $V\sim a^n$). 
\par
The combination in the square brackets effectively
measures the sub-extensive contribution to the thermodynamic
system. With this observation, one can see that the Casimir energy (\ref{38})
has indeed been appropriately defined.
Next, we will derive a more explicit expression for this Casimir
contribution.
\par
As a first step in this derivation, it is useful to
express the CFT energy density (by way of Eqs.(\ref{5},\ref{18})) 
as follows:
\bea 
\rho &=& -{ML\over a^{n+1}V_n}
\nonumber
\\
&=& {na_H^n\over 16\pi G_{n+2}a^{n+1}}\left[{a_H\over L}-{L\over a_H}
-{nL\omega^2_{n+1}Q^2\over 8(n-1)a_H^{2n-1}}\right].
\label{42}
\eea
\par
Next, we incorporate $p=\rho/n$, Eq.(\ref{34}) for $s$, Eq.(\ref{19})
for $T$ and Eq.(\ref{21}) for $\Phi$ into the above. This procedure
ultimately yields:
\be
n\left[\rho+p-\Phi\rho_Q-Ts\right]=-{2\gamma\over a^2},
\label{43}
\ee
where:
\be
\gamma\equiv {n(n-1)a_H^{n-1}\over 16\pi G a^{n-1}}.
\label{44}
\ee
\par
Recalling the  definition of the Casimir energy (\ref{38}), we have:
\be
E_C=-{2 V \gamma \over a^2}=-{n(n-1)V_n a_H^{n-1}\over 8 \pi G a}.
\label{45}
\ee
Notably, this expression  does not depend explicitly  on
the  parameters that describe    the RNdS geometry.
\par
With the above formalism, the entropy density (\ref{34}) can
be directly related to the ``Casimir quantity'' (i.e., $\gamma$).
After some tedious manipulations, we find:
\be
s^2=\left({4\pi\over n}\right)^2
\gamma \left[\rho-{1\over 2}\Phi\rho_Q+{\gamma\over
a^2}\right].
\label{46}
\ee
Significantly, this entropy formula has a Cardy-like form \cite{car},
with $\gamma$ playing the role of a ``central charge''.\footnote{In Cardy's
formalism \cite{car}, the central charge describes the multiplicity
of massless particle species. It is clear that such a quantity should be
directly related to the Casimir energy density, as we have found.}
\par
Let us now  reconsider the special moment when the brane crosses the
cosmological
horizon.
 At this coincidence point, the first Friedmann-like  equation (\ref{22})
can be shown to follow directly from Eq.(\ref{46}).
Furthermore, the second Friedmann-like equation (\ref{23}) can be
obtained when $a=a_H$ is  imposed on Eq.(\ref{43}).
Hence, we have extended the key results of Ref.\cite{sav} for the
case of a bulk RNdS spacetime.

\section{Cosmological Considerations}
\par
Before concluding, let us  examine some of the cosmological
implications of the prior results.
First, it is useful if our generalized Cardy-Verlinde
formula (\ref{46}) is re-expressed in the following form:
\be
S=\sqrt{{2\pi a\over n}S_C\left[2\left(E-{1\over 2}\Phi Q\right)
-E_C\right]},
\label{47}
\ee
where we have suitably defined the following Casimir entropy
(in analogy with Ref.\cite{ver}):
\bea
S_C&\equiv& -{2\pi a\over n}E_C
\nonumber\\
&=& {(n-1)V_n a_H^{n-1}\over 4 G}.
\label{48}
\eea
Note that $S_C$ is strictly non-negative and depends implicitly
(but not explicitly) on $M$ and $Q$  by virtue of its
dependence on $a_H$  (cf. Eq.(\ref{5})).
\par
With regard to the Casimir entropy, it is  particularly significant that:
\be
S_C=S_{BH} \quad\quad\quad at \quad a=a_H,
\label{xxx}
\ee
where $S_{BH}$ is 
 the  Bekenstein-Hawking entropy  of Eq.(\ref{31}).
Recall that the same coincidental behavior was found  between
the Casimir energy and the  ``Bekenstein-Hawking energy'' 
($E_{BH}$);
cf. Eq.(\ref{39}).
\par
It is not difficult to show that, when $a=a_H$,
 the above relation (\ref{47}) for $S$
coincides with Eq.(\ref{24}) for  the ``Hubble entropy'' ($S_H$).
To illustrate this, let us first consider the following equivalent
form of Eq.(\ref{47}):
\be
S^2=-2\left(S_B-{\bf S}_Q\right)S_C +S_C^2.
\label{49}
\ee
Here, $S_B$ is the ``Bekenstein entropy'' of Eq.(\ref{30}) and
we have further defined (in analogy to Eq.(\ref{32})):
\be
{\bf S}_Q\equiv -{2\pi a\over n}\cdot {1\over 2}\Phi Q.
\label{50}
\ee
\par
Comparing Eq.(\ref{49}) for $S$ with Eq.(\ref{29}) for $S_H$,
we clearly observe the  equivalence of these two entropies
when the brane crosses the horizon.
\par
Given the outcomes of the seminal studies \cite{ver,sav},
one might wonder if  such  enthropic coincidences
(at $a=a_H$) actually represent the saturation points of
holographic bounds.  
Before considering this, we point out a significant difference between
 dS-bulk scenarios and their  AdS analogues.  As it so happens,
the quantity $H^2 a^2$ must always be less than unity in the dS-bulk case.
 One can see this in a number of ways; for instance,
an inspection of Eq.(\ref{22}), keeping in mind
that the CFT energy density is always negative in dS 
spacetimes.\footnote{We are assuming here that $|\rho|>{1\over 2}|\phi\rho_Q|$.
This is equivalent to saying that the radiative energy dominates
over the electrostatic energy, which seems a reasonable viewpoint.}
Hence, in accordance with Verlinde's classification scheme \cite{ver},
a  dS bulk can only induce  a {\bf weakly} self-gravitating brane universe.
\par
It is interesting to note that Verlinde's ``litmus test'' for
a  weakly self-gravitating system, $H^2 a^2\leq 1$ when $S_B\leq S_{BH}$ 
\cite{ver},
translates to our  RNdS model as:
\be
S_B-S_Q\leq {1\over 2} S_{BH} \quad\quad\quad always.
\label{51}
\ee
This constraint can readily be obtained from Eq.(\ref{29}),
keeping in mind that $S_{H}^2$, $S_B$, $S_{BH}$ are strictly
non-negative (and we have implicitly assumed that $S_B>S_Q$ $^{13}$).
\par
In Ref.\cite{ver}, Verlinde conjectured a universal bound
that is based on a holographic upper limit on the degrees of freedom
of the CFT as measured by the Casimir energy.
That is,  $|E_C|\leq
|E_{BH}|$\footnote{We have included absolute-value bars to
account for the negative brane  energy that is induced by the RNdS bulk.}
or  equivalently:
\be
S_C \leq S_{BH}.
\label{52}
\ee
We will assume that this intuitive  bound continues 
to hold for the RNdS-bulk scenario 
without modification. The justification being: (i) the 
equivalence  of these entropies 
 when the brane crosses the horizon and (ii) the Casimir
energy and entropy have no explicit dependence on either $M$ or  $Q$. 
%\cite{you3}.
However, even with this conjecture, $S\leq S_H$  does {\bf not} 
necessarily hold up,
except for the case of vanishing charge; cf. Eqs.(\ref{29},\ref{49}).
(Note that ${\bf S}_Q > S_Q$ when $a_H<a$.) 
\par
We can still obtain a universal bound for the ``total'' entropy
by way of the following argument.  Again assuming that the
radiative matter dominates over the electrostatic contribution
so that $S_B > {\bf S}_Q$, we have from Eqs.(\ref{49},\ref{52}): 
\be
S < S_C\leq S_{BH}.
\label{53}
\ee
Not only does a dS black hole excite negative energy,
 it effectively lowers the entropy from that of a pure (i.e., $M=0$)
dS  state. This result seems to be a counter-intuitive outcome. 
This does, however, agree with Bousso's observation \cite{booz}:
the entropy of a pure de Sitter space serves as an upper bound for
the entropy of any asymptotically dS space.
Also  note
that a similar result has been obtained in Refs.\cite{new1,new3}.

\section{Conclusion}
\par

In the preceding paper, we have considered a brane universe in a 
Reissner-Nordstrom-de Sitter background  spacetime. 
The analysis began with the identification of thermodynamics
of the RNdS cosmological horizon. We then considered brane
dynamics and 
demonstrated that (with a suitable choice of brane tension)
the  evolution equations  take on a Friedmann-like form.
\par
Next, we applied the dS/CFT correspondence to obtain the
thermodynamic properties of  a Euclidean CFT that lives
on the brane. After which, it was explicitly shown that
the CFT thermodynamics coincides with the (generalized) 
Friedmann equations at the point when the brane crosses
the cosmological horizon. Moreover, we were able to
derive an expression for the entropy that
is readily identifiable as a generalized Cardy-Verlinde
formula \cite{car,ver}.
In this context, the Casimir energy (i.e., the sub-extensive energy
 contribution) adopts the role of the Cardy central
charge.
\par
Finally, we have  considered some of the cosmological implications of our
analysis. We found that the ``Casimir entropy'' coincides
with the Bekenstein-Hawking entropy when the brane crosses the horizon. 
On the basis of this result, we conjectured 
 an upper
bound for the Casimir entropy. That is to say, the equivalence
of entropies (when the brane and horizon coincide) is really
just a saturation of this conjectured upper bound.
We again  point out that  such a bound was previously suggested
by Verlinde \cite{ver} as a universal consequence of
the holographic principle \cite{tho,sus}.
\par
Although our results came  out clearly in support of the
proposed dS/CFT correspondence, we have  encountered
some troubling issues along the way.
These include: (i) the total entropy of the CFT  being bounded (from above)
by the Casimir contribution and (ii) the  inaccessibility of
 dS-induced  brane universes to a strongly self-gravitating regime.
Such ``exotic'' behavior 
(in comparison with  analogous AdS scenarios)
can seemingly be attributed to the negative energy density that
arises on the brane. This unusual brane property can, in turn,
be linked (via the dS/CFT duality) to dS black holes describing
 an excitation of negative energy \cite{baln,myun}.
This strange effect should probably be better
understood  before the dS/CFT correspondence can be put
on equal footing with its AdS analogue. 
Furthermore, there remains the open issue of how to incorporate
the thermodynamics of the dS black hole horizon into
the proposed duality.
\par
Suffice it to say, the dS/CFT correspondence requires
further investigation; although  considerable progress
has definitely been made.

\section{Acknowledgments}
\par
The author  would like to thank  V.P.  Frolov  for helpful
conversations.  The author would also like to thank
G. Ellis for  getting him interested in the topic of holography
(way back when).

\par\vspace*{20pt}

%\newpage

\end{document}